\documentclass[twocolumn,prb,aps,showpacs]{revtex4}
\usepackage{amssymb}

\usepackage{amsmath}
\usepackage{graphicx}
\usepackage{dcolumn}
\usepackage{bm}

\begin{document}

\preprint{Contact v.4}
\title{Lateral diffusive spin transport in layered structures}
\author{H.~Dery}\email{hdery@ucsd.edu}
\author{\L.~Cywi{\'n}ski}
\author{L.~J.~Sham}
\affiliation{Department of Physics, University of California San
Diego, La Jolla, California, 92093-0319}

\begin{abstract}
A one dimensional theory of lateral spin-polarized transport is
derived from the two dimensional flow in the vertical cross section
of a stack of ferromagnetic and paramagnetic layers. This takes into
account the influence of the lead on the lateral current underneath,
in contrast to the conventional 1D modeling by the collinear
configuration of lead/channel/lead. Our theory is convenient and
appropriate for the current in plane configuration of  an
all-metallic spintronics structure as well as for the planar
structure of a semiconductor with ferromagnetic contacts. For both
systems we predict the optimal contact width for maximal
magnetoresistance and propose an electrical measurement of the spin
diffusion length for a wide range of materials.
\end{abstract}
 \maketitle

Spintronics promises an increase in the logical expressibility of
electronic circuits and the integration of non-volatile magnetic
memory \cite{Wolf_Science00,Zutic_RMP04} . A quantitative and yet
simple theory of spin transport is essential for interpreting
experimental results and designing practical lateral devices which
are imperative in large scale integrations. The most popular
analytical approach to spin transport extends drift-diffusion
equations to include the spin degree of freedom
\cite{VanSon_PRL87,Valet_PRB93,Hershfield_PRB97,Yu_Flatte_long_PRB02}.
One-dimensional equations of this kind have successfully explained
the current-perpendicular-to-the-plane giant magnetoresistance
\cite{Valet_PRB93} and  the spin injection from a ferromagnetic
metal into a semiconductor in a vertical geometry
\cite{VanSon_PRL87,Schmidt_PRB00,Rashba_PRB00,Yu_Flatte_long_PRB02,Albrecht_PRB03}.
Theories of spin transport for current in plane focus on boundary
scattering to explain the magnetoresistive effect in ultrathin
layered system \cite{Levy_SSP94,hershfield_cip}. However, many of
the experiments involving non-magnetic metals
\cite{Johnson_Science93,Jedema_Nature01} or semiconductors
\cite{Stephens_PRL04,Crooker_Science05} are performed in the planar
geometry where the thickness of the layers exceeds the mean free
path. For these systems, application of either of the existing 1D
models is a rough approximation, unable to account for some observed
features whose interpretation required a numerical 3D treatment
\cite{Hamrle_PRB05}.
The role of the contact width in lengthening the effective path
between two leads versus the spin diffusion length was mentioned
\cite{Fert_PRB01}.  The effect of leakage from two-dimensional
electron gas into the ferromagnetic gate of finite lateral size on
the spin currents in the channel has been investigated
\cite{McGuire_PRB04}. What remains lacking is a simple theory in
which the spin injection from contacts of finite width is normal to
the spin current in the plane of the conduction channel.

In this paper, starting from a study of the two dimensional current flow in
the vertical cross section of the planar system, we derive an
analytical 1D theory of spin transport in lateral structures. The
analysis takes into account the lateral extent of the interface
between ferromagnetic (FM) and normal (N) materials.
We obtain a transparent physical picture and derive useful
analytical formulas. The most important result obtained is the
dependence of the magneto-resistance (MR) on the contact width. The
knowledge of this aspect of lateral spin-valve physics is essential
to the design of structures with high MR. In addition, our analysis
suggests a technique to obtain the spin diffusion length of a
paramagnetic material by a simple set of electrical measurements.
For compounds, such as silicon, which are inaccessible to optical
characterization of spin polarization
\cite{Stephens_PRL04,Crooker_Science05}, development of the
electrical method is of great importance.

\begin{figure}
\includegraphics[height=2.75cm,width=7cm]{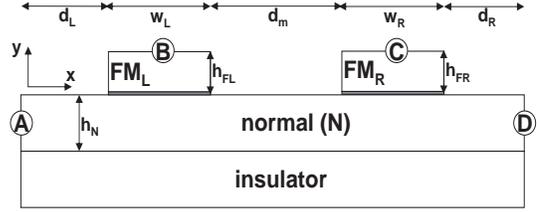}
\caption{A two-dimensional sketch of a planar system with
ferromagnetic and normal materials. A-D represent boundaries which
may be connected to the external circuit. All of the labeled lengths
enter into  our 1D transport equations. } \label{fig:scheme}
\end{figure}

Fig.~\ref{fig:scheme} depicts a typical planar system, homogeneous
in the $z$ direction. A model of one dimensional current flow from
lead B via the normal region to  lead C cannot describe the flow
under the leads because of  the simultaneous in-plane flow and the
injection or extraction from above.  Hence, we must first consider
the two-dimensional  transport governed by the spin diffusion
equation:
\begin{equation}
\nabla^2 \mu_{s}(x,y) = \frac{\mu_{s}(x,y) -
\mu_{-s}(x,y)}{(L_{s})^2} \,\, , \label{eq:diffusion}
\end{equation}
where $\mu_{s}$ is the spin dependent electrochemical potential with
$s$=$\pm$ denoting the spin species. The characteristic distance of
spin flip is $L_{s}=\sqrt{D_s\tau_{s,-s}}$, where $\tau_{s,-s}$ is
the spin-flip time and $D_s$ is the diffusion constant of the spin
component $s$. The measurable   spin diffusion length $L$, on whose
scale the splitting of $\mu$ changes, is given by
$L^{-2}$$=$$L_{s}^{-2}+L_{-s}^{-2}$. These equations hold for
ferromagnetic and normal metals \cite{Hershfield_PRB97} and also
hold for non-degenerate semiconductors with the additional condition
of a small electric field \cite{Yu_Flatte_long_PRB02}. We neglect
interfacial spin scattering and use Ohm's law across the interfaces.
Since the current is driven by the gradient of the electrochemical
potential,   the continuity of  the normal component of the spin
current across two adjacent layers $i$ and $k$  provides the
boundary conditions,
\begin{eqnarray}
 \sigma^i_{s} \Big(\widehat{n}^i\cdot\nabla\mu^i_{s}\Big)
=   G^{i,k}_s (\mu^{k}_s - \mu^i_s) =  -\sigma^k_{s}
\Big(\widehat{n}^k\cdot\nabla\mu^k_{s}\Big)  \, ,
\label{eq:boundaries}
\end{eqnarray}
where $\sigma^i_s$ and $\widehat{n}^i$ refer to layer $i$
conductivity and outward interface normal, respectively, and
$G^{i,k}_s$ is the spin-dependent barrier conductance.
Although Fig.~\ref{fig:scheme} illustrates only two layers (normal
and FM),  the equations derived below hold for  multilayers.

To reduce the essential features of the two dimensional flow to one
dimension, we define the vertical ($y$) average of $\mu_{s}$  in
each layer,
\begin{eqnarray}
\xi^{i}_{s}(x)=\frac{1}{h^i} \int_{y_0^i}^{y_1^i} dy \mu_{s}^i(x,y),
\label{eq:avg_mu}
\end{eqnarray}
with $h^i$ being the thickness of the layer between its boundaries,
$(y_0^i, y_1^i)$. For thin  layers, $ \mu_{s}^i(x,y)$ may be
replaced by its vertical average $\xi^{i}_{s}(x)$.
The conditions of validity follow the requirement that the gradient
correction along $y$ to $\mu_{s}$ in Eq.~(\ref{eq:avg_mu}) be
negligible  to  $O(h^2)$,
\begin{eqnarray} \text{(i)}~~h \ll
\sigma_s/G_s ,~~~~\text{and}~~~~\text{(ii)}~~h<L_s,
\end{eqnarray}
 with the help of Eqs.~(\ref{eq:diffusion}) and (\ref{eq:boundaries}).
Under these conditions we can derive a set of 1D equations governing
the lateral (i.e., in the layer plane) spin transport. Integrating
out the $y$ dependence in Eq.~(\ref{eq:diffusion}) and using
Eq.~(\ref{eq:boundaries}) yields,
\begin{eqnarray}
\frac{\partial \xi^{i}_{s}(x)}{\partial x^2} &=&
\frac{\xi^{i}_{s}(x) - \xi^{i}_{-s}(x)}{(L^i_{s})^2} + \frac{G^{i,i
+ 1}_s}{\sigma_s^i h^i} \Big[ \xi^i_{s}(x) - \xi^{i+1}_{s}(x)  \Big]\nonumber \\
&& + \frac{G^{i,i-1}_s}{\sigma_s^i h^i} \Big[ \xi^i_{s}(x) -
\xi^{i-1}_{s}(x) \Big] \label{eq:diffusion_avg}  \, .
\end{eqnarray}
This kinetic theory assumes that  the relevant length scales exceed
the electron mean free path. Thus, the present analysis cannot
address the quantum realm of the MR effect in ultrathin layered
systems \cite{Levy_SSP94}.

We now divide the lateral transport into regions of vertical stacks.
Consider a region with a vertical stack of $N_L$  layers. For
example, in the region of width $w_L$ covered by the left lead in
Fig.~\ref{fig:scheme}, $N_L=2$, excluding the insulator. Transport
in the layers of the stack is governed by  $2N_L$ coupled members of
Eq.~(\ref{eq:diffusion_avg}) which in the matrix form are simply,
\begin{eqnarray}
(\partial^2 / \partial x^2) \boldsymbol{\xi} = \mathbf{M} \cdot
{\boldsymbol{\xi}}\, , \label{eq:compact_form}
\end{eqnarray}
with the column vector $\boldsymbol{\xi}$ of elements  $\xi^{i}_{s}$
and the  positive definite matrix $\mathbf{M}$ of elements from
Eq.~(\ref{eq:diffusion_avg}). The solution has the form,
\begin{eqnarray}
\boldsymbol{\xi}(x) = \mathbf{1} (a + bx) + \sum_{j=1}^{N_L-1}
\mathbf{v_j}\Big( p_j e^{\lambda_j x} + q_j e^{- \lambda_j x}
\Big),  \label{eq:sol_form}
\end{eqnarray}
where $ \mathbf{1}$, a column of ones, is an eigenvector of
$\mathbf{M}$ with zero eigenvalue and $\mathbf{v_j}$ is an
eigenvector with eigenvalue $\lambda_j^2$. The connection of the
vertical stacks to the outside of the system and to each other is
maintained by the boundary conditions. Between stacks they are given
by the continuity of $\xi_{s}(x)$ and their first derivatives
(currents) through each homogeneous layer. At the outermost
boundaries (like A and D in Fig.~\ref{fig:scheme}) the conditions are
prescribed by the external  driving terms of the electrochemical
potential in the form of either a constant voltage maintained at an
electrode or an injection current (including zero) through an
interface.
Applying currents/voltages to B and C interfaces results in the
presence of inhomogenous terms in Eq.~(\ref{eq:compact_form}).
These boundary conditions provide a
unique set of solutions for the parameters $a,b,p_j,q_j$ in
Eq.~(\ref{eq:sol_form}). This method of solution is a considerable
simplification compared to solving the 2D spin diffusion equation
for the same planar structure, and yet it includes the influence of
the vertical layers on the lateral spin current flow. In the
following, we will show the versatility of this simplified approach
as well as testing its validity against the 2D solution.

The first illustrative example involves a semiconductor in the
normal layer of the structure in Fig.~\ref{fig:scheme}. A and D are
disconnected from outside while B and C are voltage biased. The spin
diffusion length in the normal layer is $L_N$. We divide the normal
channel into sections belonging to 5 vertical stacks as shown in
Fig.~\ref{fig:scheme}. In the middle section of the channel, the
solution is  \cite{Hershfield_PRB97,Yu_Flatte_long_PRB02},
\begin{eqnarray}
\xi_{\pm}(x)= (e/\sigma_{N})Jx+\mu\pm( p_{c}e^{x/L_{_N}} +
q_{c}e^{-x/L_{_N}}), \label{eq:mid_channel}
\end{eqnarray}
where $J,\mu,p_{c}$ and $q_{c}$ are constants to be determined by
the boundary conditions. The total current flowing between the two
leads is $I=J h_{_N}$ per unit  length of the structure in the $z$
direction. For the two sections outside the footprint defined by the
two leads, we introduce the notion of the ``open'' versus
``confined'' geometry depending on whether $d_L,d_R \gg $~or~$\ll
L_{N}$. In both geometries the pattern of the net charge current
occurs only between the leads B and C. This means that the outside
sections lack the linear term of Eq.~(\ref{eq:mid_channel}).
Additionally the ``open'' geometry includes only an exponentially
decaying solution away from the closer lead, and in contrast with
the confined geometry, spin polarization extends noticeably outside
of the footprint between B and C, thus reducing the spin
accumulation densities. In each of the vertical stacks containing a
lead, due to vastly different conductivities of the semiconductor
and ferromagnetic metal we can decouple the equations for $\xi_{s}$
in the N semiconductor and in the FM metal and solve for the
eigenvalues of $\mathbf{M}$ in the former,
\begin{eqnarray}
\lambda^2_{(s,c)} &=& [\alpha+ 1 \pm  \sqrt{1+ \beta^2}]/(2L_N^2),
 \nonumber \\
( \alpha,\beta) &=& 2L_N^2(G_+\pm G_-)/(\sigma_{_N}h_{_N}),
\label{eq:cont_def}
\end{eqnarray}
where the first of each pair of symbols $(s,c)$ or $(\alpha,\beta)$
takes the upper sign. The electrochemical potential in the FM is
practically a constant ($\xi^{FM}$$=$$-eV$), and in the
semiconductor layer under the biased lead is
\begin{eqnarray}
\xi_{\pm}(x) &=& -eV + (1
 \pm  \lambda)\Big[ p e^{\lambda_{s} x}+
q e^{-\lambda_{s} x}  \Big] \nonumber \\ && \,\,\, + (\lambda \mp
1)\Big[ r e^{\lambda_c x}+ s  e^{-\lambda_c x} \Big] ,
\label{eq:cont_final}
\end{eqnarray}
where $p$, $q$, $r$ and $s$ are constants to be determined by
boundary conditions and $\lambda =  \cot[\frac{1}{2}
\tan^{-1}\beta]$. Consider the case that spin polarization is robust
so that $\alpha$ and $\beta$ are comparable. If $\alpha \ll 1$, then
$\lambda_c \ll 1/L_N$ and  $\lambda_s \sim 1/L_N$. The $s$-mode is
limited by the spin diffusion constant, and it corresponds to spin
accumulation ($\lambda$$\gg$$1$ in this case). If $\alpha \gg 1$,
then both eigenvalues are nearly independent of $L_N$, and neither
of the eigenvectors is a pure spin mode $\lambda$$\simeq$$1$: the
inhomogeneity of injection dominates the spatial dependence.

To illustrate the effect of the contact width on MR, we use the
experimentally verified barrier parameters of a Fe/GaAs system at
300K \cite{Hanbicki_APL02,Hanbicki_APL03}. The tunneling
conductances are of the order of $10^{2}$-$10^{3}\,
\Omega^{-1}$cm$^{-2}$ with the ratio of spin-up to spin-down
conductance $G_{+}/G_{-}\simeq 2$. The N channel doping is
$n_0$$=$$4 \times 10^{15}$ cm$^{-3}$ and the spin relaxation time is
$\tau_{s}$$=$$80$ ps \cite{Optical_Orientation} which corresponds to
$L_{_N}$$\simeq$1$~\mu$m. The thickness of the N channel is
$h_{_N}$$=$$100$ nm, and the inner channel length is $d_m$$=$$200$
nm.

\begin{figure}
\includegraphics[height=5cm,width=8cm]{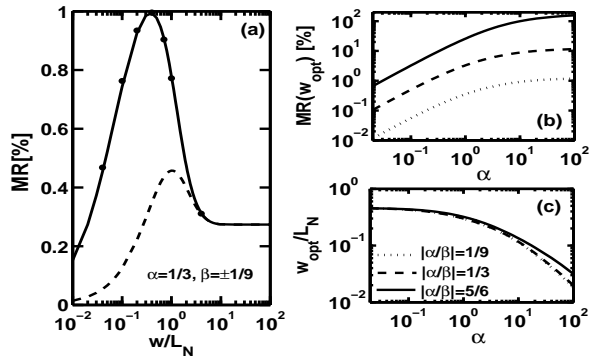}
\caption{(a) Magneto-resistance effect versus the contact width of a
GaAs channel at 300K. The solid (dashed) line denotes a confined
(open) structure. The dots are the results of a 2D numerical computation in the confined geometry. (b) Magneto-resistive effect versus
$\alpha$ for three cases of spin selectivity, $\beta/\alpha$ in the
confined geometry, calculated for the optimal contact width
corresponding to the same $\alpha$ value shown in (c).
\label{fig:MR}}
\end{figure}
The MR effect is defined as $($$I^P$$-$$I^{AP}$$)$$/$$I^{P}$ with
$I^P$ ($I^{AP}$) being the total current when the FM contacts are
magnetized in parallel (antiparallel) directions. The  calculated MR
is shown in Fig.~\ref{fig:MR}(a) as a function of the contact width,
where for simplicity we have used $w$$=$$w_L$$=$$w_R$. In the
confined geometry, $d_L$$=$$d_R$=$0$, the effective 1D method (solid
line) is compared with the results of 2D numerical calculation
(dots), showing excellent agreement. We note the deleterious effect
of the open geometry (dashed line) compared with the confined
geometry (solid line) because of the weaker spin accumulation in the
semiconductor channel which produces the MR effect \cite{Dery_cm}.

Note the existence of the optimal contact width for the maximum MR
effect which arises out of the balance between spin injection and
spin relaxation in the channel. The spin accumulation in the normal
conduction channel is built up from injection along the width of the
ferromagnetic contact. When the width is small, inadequate build-up
results in small spin injection and spin accumulation as a
percentage of the equilibrium carrier density. Consequently the MR
effect is small.  However, when the contact width is too large, its
resistance becomes very small, the MR effect in the N channel is
again negligible as in the conventional one-dimensional model
\cite{Rashba_PRB00}.   Finally, when the contact width $w$ exceeds
the spin diffusion length $L_{N}$, the build-up from vertical
injection along the width of the contact beyond $L_{N}$ becomes
ineffective, the spin injection and the MR effect approach
asymptotic values. The difference between the open and confined
structures is removed, as can be seen from the merging of the solid
and dashed lines in Fig.~\ref{fig:MR}(a). The magnitude of MR at the
optimal contact width depends also on the spin selectivity of the
barriers, $\beta$  in Eq.~(\ref{eq:cont_def}), as illustrated in
Fig.~\ref{fig:MR}(b). Fig.~\ref{fig:MR}(c) shows a much weaker
dependence on the spin selectivity of the optimal contact width in
units of the spin diffusion length ($w_{opt}/L_{_N}$). The ratio
depends only on $\alpha$ as defined in Eq.~(\ref{eq:cont_def}). For
$\alpha\gg 1$, $w_{opt}/L_{_N} \simeq 2/\alpha $. For  $\alpha\ll
1$, the optimal contact width $w_{opt}/L_{_N}$ is approximately 6/5
or 3/8 for the open or confined geometry, respectively. These
results enable the extraction of the spin diffusion length of a test
material by measuring the MR of several structures with different
contact widths from the same growth.

To demonstrate the use of our method for the all-metallic system, we
turn to a second example. In the planar structure of
Fig.~\ref{fig:scheme}, a constant current is injected from region A
into a 10 nm paramagnetic metal layer. The structural parameters in
terms of the spin diffusion length of the N layer are
$0.2$$d_L$$=$$0.2$$d_R$$=$$2$$d_M$$=$$L_{N}$. The FM layers B and C
are of the same thickness and serve as floating contacts. We use the
Al/Co system parameters at 300K where ratio of spin-up to spin-down
conductance is $G_{+}$$/$$G_{-}$$=$$2$ \cite{Vouille_PRB99} and
their sum is $1.5\times 10^9 \Omega^{-1}\text{cm}^{-2}$.
The normal layer intrinsic
parameters are $\sigma_{_N}$$=$$3$$\times$$10^5
\Omega^{-1}\text{cm}^{-1}$,
$L_{_N}$$=$$0.35$$~\mu$m. In the Co layers the majority channel
conductivity and spin diffusion lengths are, respectively, $0.1
\sigma_{_N}$ and 12 nm. The respective values of the Co minority
channel are $0.04 \sigma_{_N}$ and 20 nm. For ease of toggling
between $P$ and $AP$ configurations we use a width ratio of 2
between left and right Co regions. For the FM/N regions the
solutions are given by Eq.~(\ref{eq:sol_form}). In the middle N
channel the solution of $\xi_{s}(x)$ is given by
Eq.~(\ref{eq:mid_channel}) while at the outer channels the form of
$\xi_{s}(x)$ lacks the diverging exponential.

\begin{figure}
\includegraphics[height=5cm,width=8cm]{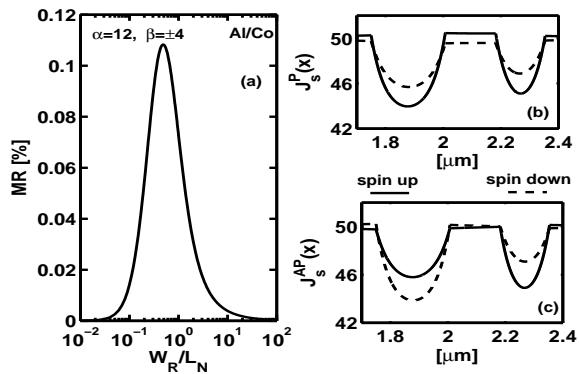}
\caption{(a) The magneto-resistive effect versus the right contact
width in Al/Co system at 300K. (b) The current density profile in
the Al layer for parallel magnetic configuration at the optimal
contact width. Solid (dash) lines denote spin-up (down) current
components. The total current is constant where in the dip regions
along the Al/Co interfaces currents propagate in both layers. (c)
Same as (b) for the anti-parallel magnetic configuration.}
\label{fig:metal}
\end{figure}

For each magnetic configuration ($P$ or $AP$) we calculate the
voltage drop $V_\text{BC}$ with the electrochemical potentials
averaged along each Co layer. The MR is defined as
$|$$1$$-$$V^{P}_\text{BC}$$/$$V^{AP}_\text{BC}$$|$.
Fig.~\ref{fig:metal}(a) shows the MR effect. It peaks at the right
contact width of $w_R\sim0.5 L_{_N}$.
The optimal contact width mostly on $\alpha$, insensitive to varying
the spin selectivity, the middle N channel length, or the outer N
channel lengths (not shown). Fig.~\ref{fig:metal}(b) and (c) show,
respectively, the current densities in the normal layer for parallel
and antiparallel magnetic configuration at the optimal contact
width. The solid (dashed) lines depict spin up and spin down current
components along the x direction. Despite the small MR effect, its
observation is well within existing experimental abilities. Thus a
similar analysis to the previous FM/SC example leads to an
electrical method to measure the spin diffusion length of the normal
material. However, the extraction of $L_N$ requires that it greatly
exceeds the FM spin diffusion length.

In summary, we have presented an effective 1D theory which describes
spin dependent transport beneath and between ferromagnetic contacts
in the lateral geometry. Under the conditions of thin layers and
non-ohmic interfaces, the derived set of coupled linear equations
governing the lateral diffusive spin currents is almost as accurate
as the 2D spin diffusion equation but much simpler to use. This
method retains the important role of the contact width, predicting
an optimal contact width for the maximal magneto-resistive effect in
the lateral spin-valve. We have also quantified the effect of
``open'' channels, where spins can diffuse away from the active
region underneath the FM contacts, in contrast with the ``confined''
structures, where the spin accumulation is kept between the two
contacts.
Our results provide crucial guidelines for design of lateral
spin-valve type devices. They also yield a method of extracting the
spin-diffusion length in an all-electrical measurement. At room
temperature where the spin diffusion length lies in the sub-micron
region, this method does not have the difficulty of wavelength
resolution in the optical techniques.

This work is supported by NSF DMR-0325599.

\end{document}